\documentclass{SolarPhysics}
\usepackage[optionalrh]{spr-sola-addons} 
\usepackage{graphicx}        
\usepackage{color}           
\usepackage{url}             
\usepackage{amstext,amsbsy,amsopn,amsgen,amssymb}

\newcommand\vecb{\boldsymbol}
\newcommand\ic{{\rm i}}

\def\omref{\omega_{\rm ref}}
\def\gl{\gamma_l}
\def\gs{\gamma_s}
\def\gls{\gamma_{l'}}
\def\pl{\partial}
\def\rsun{r_\odot}

\begin{document}

\begin{article}

\begin{opening}

\title{Meridional Circulation and Global Solar Oscillations}

\author{M. Roth$^{1}$\sep
        M. Stix$^{2}$\sep
       }
\runningauthor{Roth \& Stix}
\runningtitle{Meridional circulation and p-modes}

   \institute{$^{1}$ Max-Planck-Institut f\"ur Sonnensystemforschung,
   Katlenburg-Lindau, Germany
                     email: \url{roth@mps.mpg.de}\\
              $^{2}$ Kiepenheuer-Institut f\"ur Sonnensphysik, Freiburg,
              Germany
                     email: \url{stix@kis.uni-freiburg.de} \\
             }

\begin{abstract}
We investigate the influence of large-scale meridional circulation on solar p-modes by
quasi-degenerate perturbation theory, as proposed by~\inlinecite{lavely92}. As an input
flow we use various models of stationary meridional circulation obeying the continuity
equation.
This flow perturbs the eigenmodes of an equilibrium model of the Sun.
We derive the signatures of the meridional circulation in the frequency multiplets of solar
p-modes. In most cases the meridional circulation leads to negative average frequency shifts of the multiplets.
Further possible observable effects are briefly discussed.
\end{abstract}
\end{opening}

\section{Introduction}
Meridional circulation is a large-scale flow observed on both
hemispheres of the solar surface~\cite{duvall79,hathaway96,komm93}. Its
predominant direction is from the equator to the poles, and its amplitude
is of the order of 15\thinspace m/s. As mass does not accumulate in the
polar regions a return flow from the poles to the equator is suspected
deeper within the solar interior. The top half of the convection zone
contains approximately 0.25\% of the solar mass, the mass of the bottom
half is approximately five times larger. Consequently, a poleward flow of
10\thinspace m/s in the top half of the convection zone could be
compensated by an equatorward flow of 2\thinspace m/s in the lower half.
The transport of magnetic flux from mid to low latitudes by such a flow
at the bottom of the convection zone would last approximately 10\thinspace
years, which is close to the period of the solar magnetic cycle.

In addition to magnetic flux, the meridional flow also transports angular
momentum. Indeed, the circulation played a key role in early theories
of the solar non-uniform rotation as well as of the magnetic
cycle~\cite{bjerknes26,kippenhahn63}. More recently, differential rotation
has been explained in mean-field models as a consequence of the Reynolds
stresses~\cite{ruediger80,kueker01,kueker05},
and in three-dimensional numerical models by the influence of the Coriolis
force on global convection~\cite{miesch00,miesch06}. Nevertheless,
meridional circulation occurs as well in these models, and in solar-cycle
models it has regained popularity, since the mean-field latitude migration
along the surfaces of isorotation that occurs in the traditional
$\alpha\Omega$ dynamo~\cite{parker55} does not seem to suffice. The effect
of the circulation on the butterfly diagram had been demonstrated by
~\inlinecite{roberts72}; it is considered to be essential in more recent
versions of the $\alpha\Omega$ dynamo, which therefore have been termed
`flux-transport dynamos'~\cite{choudhuri95,dikpati99,nandy02,rempel06a,rempel06b}.

Local helioseismology has investigated the strength of the meridional
flow in the solar interior. By means of ring diagram analysis
\cite{hill88} \inlinecite{haber02} inverted data for the circulation in a
15\thinspace Mm deep region below the solar surface. In data from 1998
they found a flow emerging at high northern latitudes with equatorward orientation.
This was interpreted as an evolving second cell of
circulation. Further studies on the evolution of the flow either by
ring-diagrams and by time-distance helioseismology
(e.g.~\opencite{zhao04},~\opencite{zaatri06}) show predominantly a polward flow with a strong variability in
the outer 15\thinspace Mm of the Sun. The velocity reaches 40\thinspace m/s.
These findings were interpreted as the upper parts of meridional
circulation cells in the two hemispheres.

Theoretically, the influence of global-scale stationary flows on solar
p-modes was studied in detail by quasi-degenerate perturbation
theory~\cite{lavely92}. These studies were successfully used to solve the
forward problem for the influence of differential rotation on the p-modes.
The results lead to an improved inversion method for determining the
radial dependence of the differential rotation~\cite{ritzwoller91}.
Following~\inlinecite{lavely92}, \inlinecite{roth99} studied
the influence of large-scale sectoral poloidal flow components that could
be related with giant convection cells. They found that such flows yield
additional frequency shifts that can only be described with quasi-degenerate
perturbation theory, as these frequency shifts are effects of higher order.
In a sequel study~\inlinecite{roth02} were able to show that sectoral poloidal
flows could only be found with the current inversion methods of global
helioseismology as long as they exceed an amplitude of 10\thinspace m/s.
The meridional flow was found to be not detectable by the current inversion methods as the
frequency splittings are fitted by an incomplete set of basis functions, which are
tailored to measure only zonal toroidal flows.
However, no detailed study on the effect of the meridional circulation on the
oscillation frequencies was given, and few other attempts to derive observable
signatures of the meridional flow in global helioseismology data
exist~\cite{woodard00}.

In this contribution we concentrate on a theoretical study of the influence
of the meridional circulation on the solar p-mode frequencies and describe
a possible observable effect. This effect is significantly smaller than
the frequency splitting caused by solar differential rotation. But as time
series of global oscillation data exist that cover more than 10 years, the
necessary frequency resolution might be available to detect it. The
advantage of studying the meridional circulation by global helioseismic
techniques is a possible inference of information from greater depths.

\section{Frequency shifts caused by meridional circulation}
The effect of the meridional circulation on solar oscillations shall be investigated
by solving the forward problem. We use quasi-degenerate perturbation
theory as proposed by~\inlinecite{lavely92} to calculate shifts of the oscillation
frequencies. A short outline of the mathematics is given in the follwing.

As described by~\inlinecite{lavely92} and~\inlinecite{roth99} we consider a spherically symmetric
equilibrium model of the Sun. For this purpose ``Model S''
from~\inlinecite{jcd96} is used. The solar p-modes are adiabatic
eigen-oscillations $\vecb{\xi}_k$ of small amplitude, where $k$ stands for
the three indices harmonic degree $l$, azimuthal order $m$ and
radial order $n$. We calculate the eigenfrequencies and eigenmodes
numerically with the ADIPACK code~\cite{jcd91}.

The equation governing the eigenmodes is
\begin{equation}
\mathcal{L}_0\vecb{\xi}_k =-\rho_0\omega_k^2\vecb{\xi}_k\ ,
\label{eq1}
\end{equation}
where $\rho_0$ is the density and $\omega_k$ the oscillation frequency. The
operator $\mathcal{L}_0$ acts on the eigenoscillation by
\begin{equation}
\mathcal{L}_0\vecb{\xi}=-\nabla P'+\rho_0\vecb{g'}+\rho'\vecb{g_0}\ ,
\label{eq2}
\end{equation}
where the primed quantities are the Eulerian variations of
pressure, gravitational acceleration, and density caused by an oscillation mode
(e.g.,~\opencite{stix02},~\opencite{unno89}).

Describing the disturbing effect of the meridional circulation we replace in
Eq.~(\ref{eq1}) the operator $\mathcal{L}_0$, the squared eigenfrequency
$\omega_k^2$, and the eigenfunction $\vecb{\xi}_k$ with
\begin{eqnarray}
\mathcal{L}_0&\rightarrow& \mathcal{L}_0+ \mathcal{L}_1\ ,\nonumber\\
\omega_k^2&\rightarrow& \tilde{\omega}_j^2\ , \label{eq3}\\
\vecb{\xi}_k&\rightarrow& \tilde{\vecb{\xi}}_j\ ,\nonumber
\end{eqnarray}
where the perturbation operator $\mathcal{L}_1$ is defined in terms of the
meridional circulation $\vecb{u}_0$ and acts on the eigenmodes as
\begin{equation}
\mathcal{L}_1(\vecb{\xi}_k)= -2\ic \omref\rho_0 (\vecb{u}_0\cdot\nabla)
                                              \vecb{\xi}_k\ .
\label{eq4}
\end{equation}
The perturbed eigenfunctions are expressed in terms of the unperturbed normal
modes of the equilibrium model
\begin{equation}
\tilde{\vecb{\xi}}_j  = \sum_{k\in\mathcal{K}} a_k^j\vecb{\xi}_k\ ,
\label{eq5}
\end{equation}
where the index $k$ and the subscript $j$ are elements of the subspace
$\mathcal{K}$ that is spanned by the eigenfunctions which are
quasi-degenerate. We identify $j$ with that $k$ for which the expansion
coefficients $a_k^j$ are maximal in magnitude. Following~\inlinecite{lavely92} we
use
\begin{equation}
\tilde{\omega}^2=\omref^2+ \lambda\
\label{eq6}
\end{equation}
for the perturbed frequency. The reference frequency $\omref$ should be
chosen in the vicinity of the eigenfrequencies of the modes in the subspace
$\mathcal{K}$.

According to~\inlinecite{lavely92} the problem can be turned into an algebraic
eigenvalue problem by substituting (\ref{eq5}) and (\ref{eq6}) into the
perturbed eigenvalue problem, and making use of the orthogonality of the
eigenfunctions
\begin{equation}
\sum_{k\in{\mathcal{K}}} a_k^j Z_{k'k} = \sum_{k\in{\mathcal{K}}} a_k^j
\lambda_j \delta_{k'k} \qquad {\rm for}\ k'\in{\mathcal{K}}\ .
\label{eq7}
\end{equation}
The elements of the matrix $\mathbf{Z}$ are given by
\begin{equation}
Z_{k'k}= {1\over N_{k'}}\left\{ H_{n'n,l'l}^{m'm}
         -(\omref^2 - \omega_k^2) N_k \delta_{k'k}\right\}\ , \label{eq8}
\end{equation}
with
\begin{equation}
H_{n'n,l'l}^{m'm}= -\int \vecb{\xi}_{k'}^* \cdot
{\mathcal{L}}_1\vecb{\xi}_k\,d^3r\ .
\label{eq9}
\end{equation}
The normalization of the eigenfunctions is given by
\begin{equation}
N_k = \int_0^{\rsun}\rho_0 [\xi_r^2 + l(l+1)\xi_h^2] r^2\,dr\ .
\label{eq10}
\end{equation}

\subsection{The model of the  meridional circulation}

We use a linear superposition of Legendre polynomials $P_s(\cos\theta)$ with
degree $s$ for our representation for the meridional velocity field
\begin{equation}
\vecb{u}_0 = \sum_{s} u_s (r) c_s P_s(\cos\theta) \vecb{e}_r +
              v_s(r) \nabla_h c_s P_s(\cos\theta)\ ,
\label{eq11}
\end{equation}
where $\nabla_h$ is a horizontal gradient and $\theta$ is the colatitude. The
factor $c_s$ is used for normalization of the Legendre polynomials and is
determind by $c_s^2=(2s+1)/4\pi$. The expansion coefficients $u_s(r)$ and
$v_s(r)$ are functions of the radial coordinate $r$, and represent the radial
and horizontal strength of the flow component with degree $s$. In this model
the meridional velocity field is a zonal poloidal flow which is independent
of longitude.

The meridional circulation shall be free of divergence, i.e.
$\nabla\cdot(\rho_0\vecb{u}_0) = 0$, which enables to express $v_s$ in terms
of $u_s$,
\begin{equation}
\rho_0 r s(s+1) v_s = \partial_r(r^2\rho_0 u_s) \ .
\label{eq12}
\end{equation}

We adopt a simple model for the depth dependence of $u_s(r)$, which is
given by
\begin{eqnarray}
u_s(r) &=& A \sin(n_c \pi \frac{r-r_b}{R_\odot-r_b})\quad {\rm for}\ r_b\le r\le
R_\odot\ ,\nonumber\\
u_s(r)&=& 0 \quad\mbox{otherwise}\ .
\label{eq13}
\end{eqnarray}
The return flow of the meridional flow is closed at the radius given by the
parameter $r_b$, which we set equal to the bottom of the convection zone
$r_b=0.713 R_\odot$ \cite{basu97}. The number of circulation cells in depth is defined by the parameter $n_c$.
We select the amplitude $A$ such that the
horizontal flow component $v_s$ at the solar surface has a maximum
amplitude of 15 m/s.

\begin{figure}
\includegraphics[width=4cm]{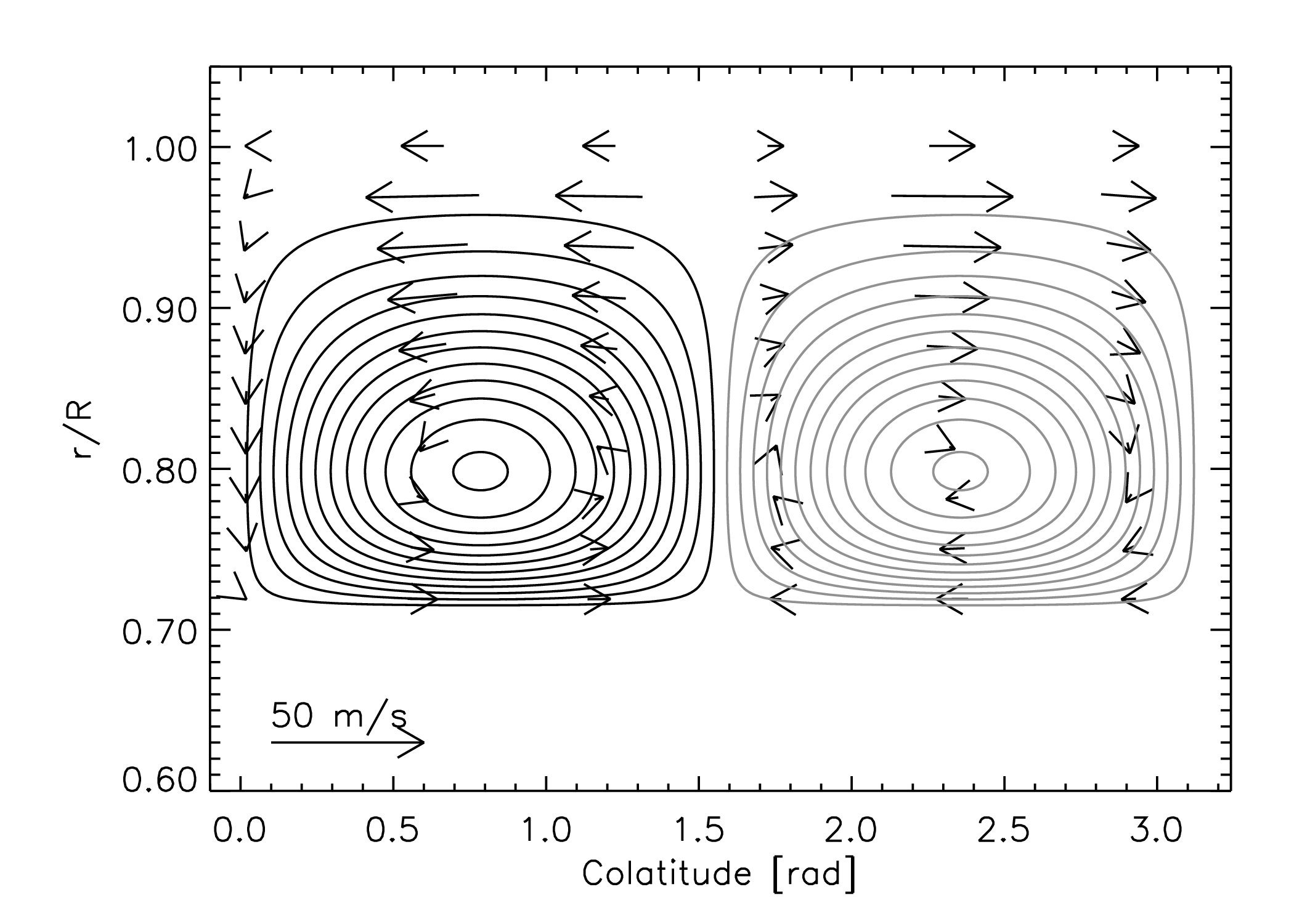}
\includegraphics[width=4cm]{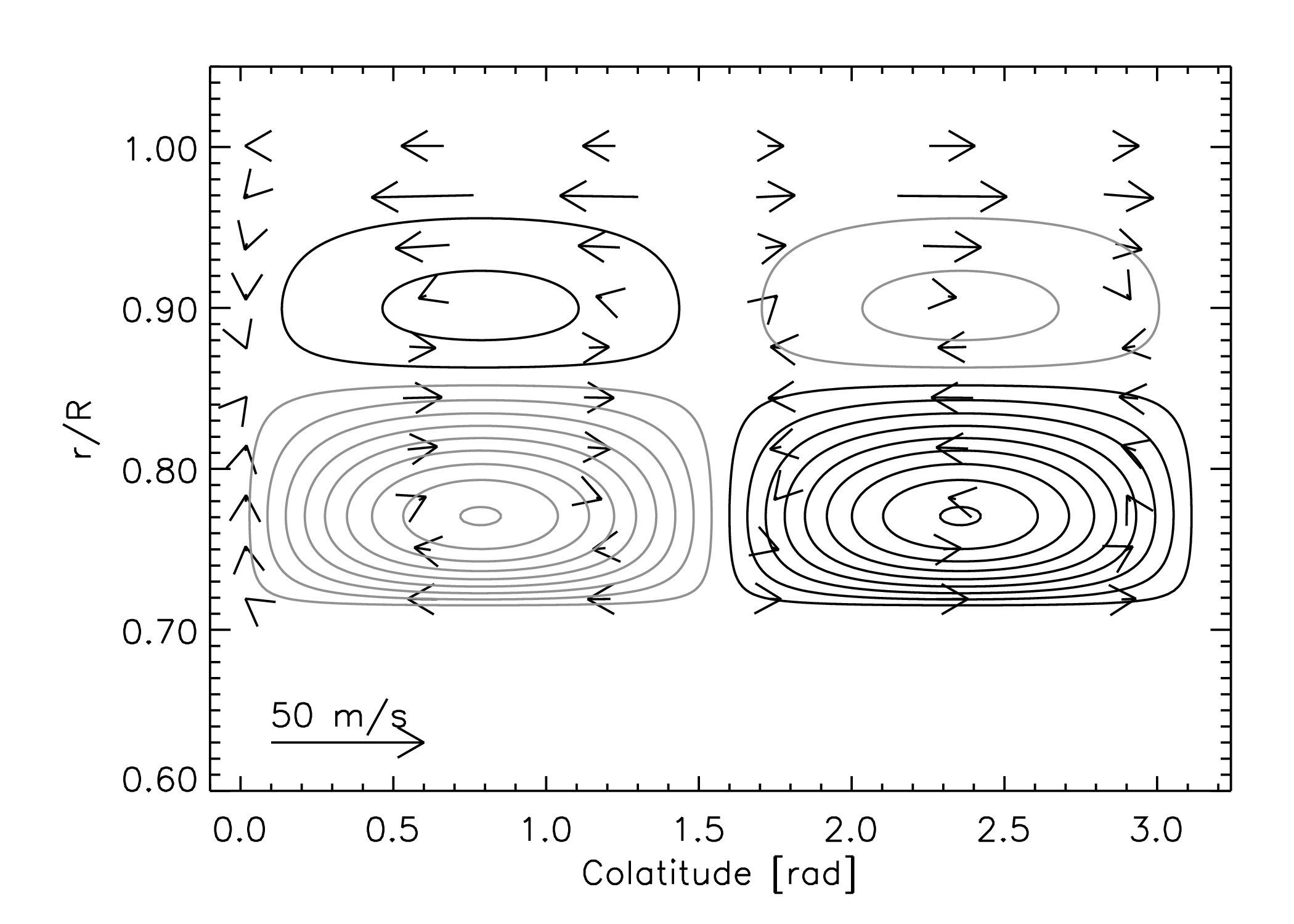}
\includegraphics[width=4cm]{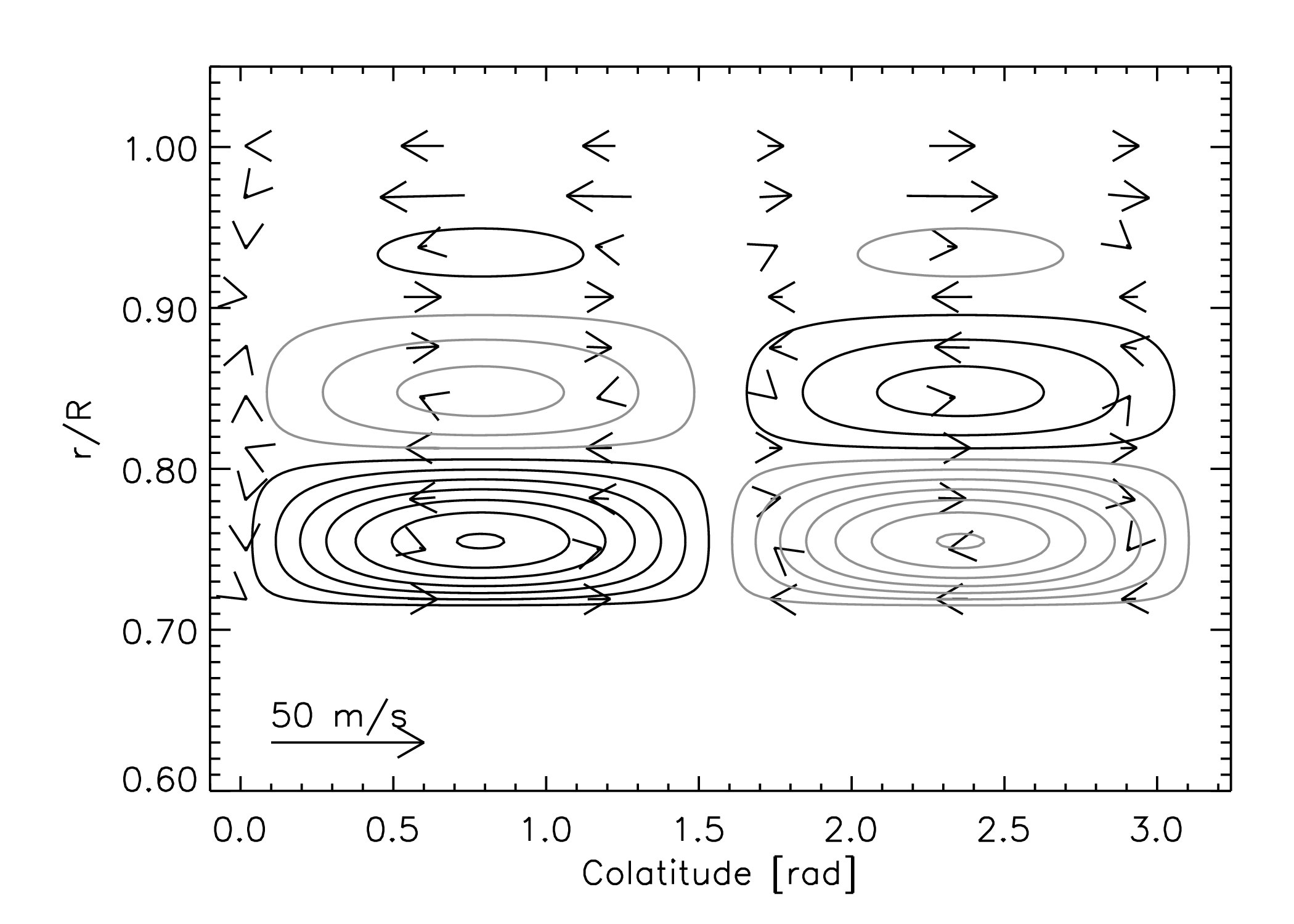}

\includegraphics[width=4cm]{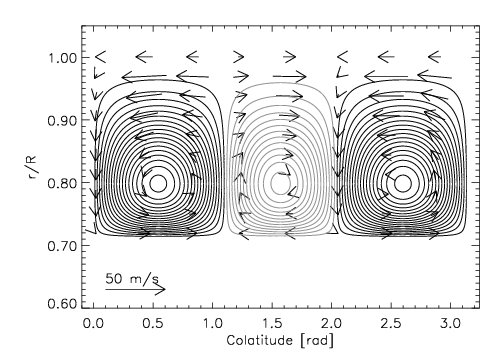}
\includegraphics[width=4cm]{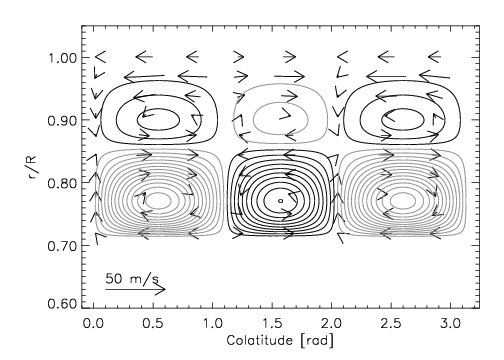}
\includegraphics[width=4cm]{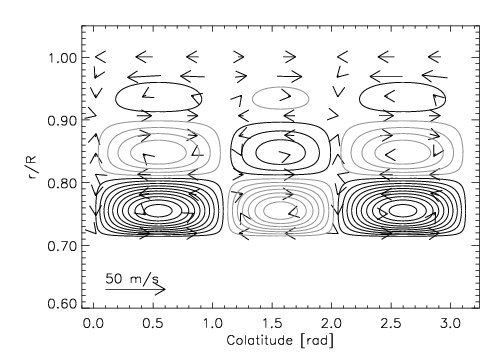}

\includegraphics[width=4cm]{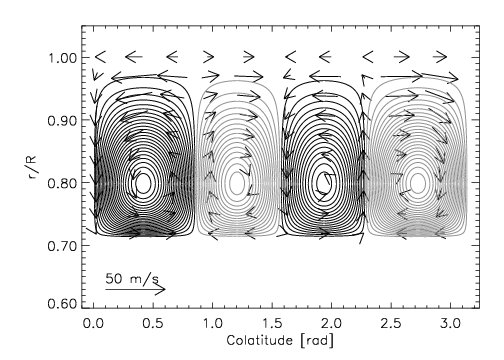}
\includegraphics[width=4cm]{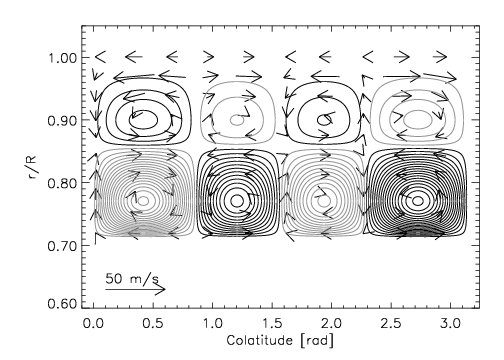}
\includegraphics[width=4cm]{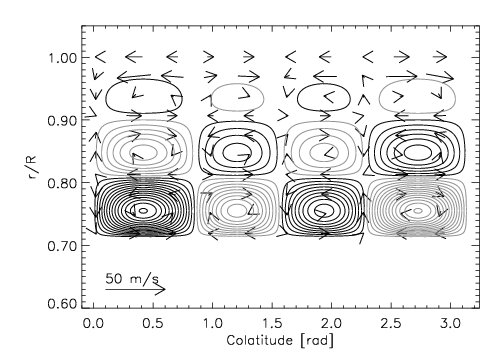}

\includegraphics[width=4cm]{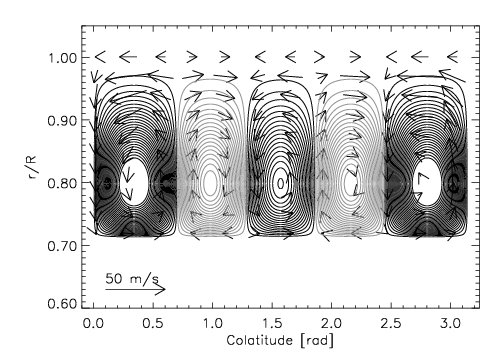}
\includegraphics[width=4cm]{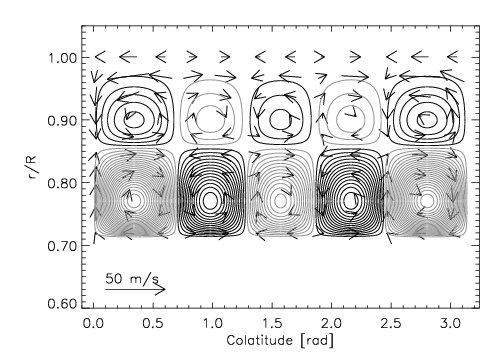}
\includegraphics[width=4cm]{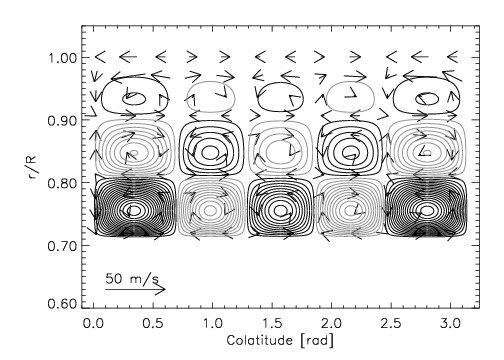}

\includegraphics[width=4cm]{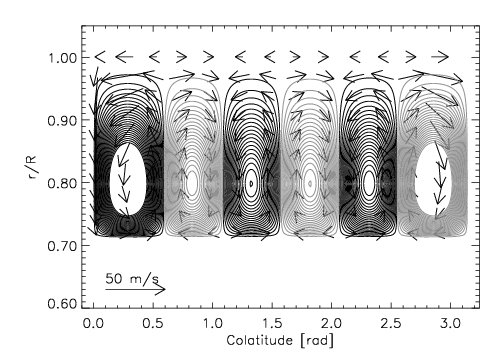}
\includegraphics[width=4cm]{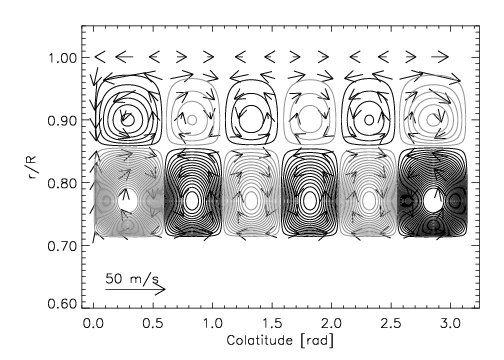}
\includegraphics[width=4cm]{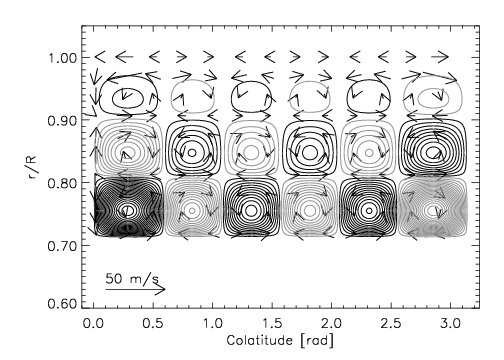}
\caption{Streamlines for the flows $\rho\vecb{u}_0$ of the 15 meridional
circulation models as functions of solar
colatitude and radius. Black indicates counterclockwise flow, grey clockwise.
The arrows indicate the flow velocity at some positions.
The number of cells in horizontal direction changes
from top to bottom between 2 and 6. The number of cells in depth changes from
left to right between 1 and 3.} \label{fig1}
\end{figure}

Below we present results of the effects on the solar p-mode frequencies
of 15 various models of the meridional circulation. We vary $s$ between 2
and 6 and $n_c$ between 1 and 3. The higher degrees $s$ represent more than
one cell in each hemisphere; such multi-cell circulation occurs in numerical
simulations~\cite{brun02,miesch06}, and has been observed
by means of sunspot tracing on the solar surface~\cite{tuominen83,woehl01}.
The models with odd $s$ have an equator-crossing
flow; we include these as such asymmetric flows have been found on the Sun
as well~\cite{haber00,zhao04}. Figure~\ref{fig1} illustrates
the 15 models.

Due to Eq.~(\ref{eq12}) models with higher degree $s$ show stronger radial
flows in the down- and upflow channels whereas the horizontal flow strength is
the same in all models.

\subsection{Theoretical frequency shifts}

We evaluate the integration of (\ref{eq9}) by using the
model of the meridional circulation (\ref{eq11}) and (\ref{eq12}).
The integral over longitude does not vanish only if $m'=m$.
The result is
\begin{eqnarray}
H_{n'n,l'l}(m) &=& 8\ic\omref\pi (-1)^m \gl\gls\nonumber \\
&&\times\sum_s \gs
\int\limits_0^{R_\odot} u_s \left[R_s
-\partial_r\left(\frac{H_s}{s(s+1)}\right)\right]\rho_0 r^2\, dr
\left(\begin{array}{ccc}s&l&l'\\0&m&-m\end{array}\right)\ .
\label{eq14}
\end{eqnarray}
The integral kernels $R_s(r)$ and $H_s(r)$ are the equivalents of the more
general poloidal flow kernels given in \inlinecite{lavely92}
\begin{eqnarray}
R_s(r) &=&\frac{1}{4}\left(\xi'_r\frac{\pl\xi_r}{dr} -
\frac{\pl\xi'_r}{dr}\xi_r
\right)(1+(-1)^{s+l+l'})\left(\begin{array}{ccc}s&l&l'\\0&0&0\end{array}\right) \nonumber\\&& +\frac{1}{2}
\left(\xi'_h\frac{\pl\xi_h}{dr} -\frac{\pl\xi'_h}{dr}\xi_h
\right)(1+(-1)^{s+l+l'})\Omega_l\Omega_{l'}\left(\begin{array}{ccc}s&l&l'\\0&1&-1\end{array}\right)
\ ,\nonumber \\
H_s(r)&=&\frac{1}{2}[l(l+1)-l'(l'+1)] (1+(-1)^{s+l+l'})\nonumber\\&&\times
\left[\xi_r\xi'_r \left(\begin{array}{ccc}s&l&l'\\0&0&0\end{array}\right) -
\xi'_h\xi_h\Omega_l\Omega_{l'}\left(\begin{array}{ccc}s&l&l'\\0&1&-1\end{array}\right)
\right]\nonumber\\ &&
-\xi'_h\xi_r
(1+(-1)^{s+l+l'})\Omega_{l'}\Omega_s\left(\begin{array}{ccc}s&l&l'\\1&0&-1\end{array}\right)
  \nonumber\\ &&
 +\xi'_r\xi_h (1+(-1)^{s+l+l'}) \Omega_l\Omega_s
\left(\begin{array}{ccc}s&l&l'\\1&-1&0\end{array}\right)\ .
\label{eq15}
\end{eqnarray}
where $\gamma_x = \sqrt{(2x+1)/4\pi}$ and $\Omega_x=\sqrt{x(x+1)/2}$. The
$3\times 2$ arrays are Wigner-$3j$ symbols that result from the coupling of
the angular momenta of the oscillations and the flow
involved~\cite{edmonds60}. Due to this coupling of angular momenta the matrix
elment $H_{n'n,l'l}(m)$ is non-vanishing if certain selection rules are
fulfilled. The first rule arises from properties of the Wigner-$3j$ symbols
which vanish exept when the harmonic degrees $l$, $l'$, and $s$ satisfy a
triangular condition. The second rule follows from (\ref{eq15}); the sum of
the degrees must be even otherwise the factor $(1+(-1)^{s+l+l'})$ in $R_s$
and $H_s$ vanishes,
\begin{equation}
s + l + l' \equiv 0 \quad {\rm mod\ } 2\ .
\label{eq16}
\end{equation}

We can simplify Eq.~(\ref{eq14}) by defining a poloidal flow kernel $K_s(r)$
\begin{equation}
H_{n'n,l'l}(m)=:8\ic\omref\pi (-1)^m \gl\gls \sum_s \gs
\int\limits_0^{R_\odot} u_s K_s(r)\rho_0 r^2 \, dr
\left(\begin{array}{ccc}s&l&l'\\0&m&-m\end{array}\right)\ .
\label{eq17}
\end{equation}

According to this result the meridional circulation leads to an effective
shift of the eigenvalues, i.e. the mode frequencies, if the selection rules
are fulfilled. These shifts are given by the eigenvalues $\lambda$ of the
matrix $\vecb{Z}$, or in more detail, the corrected mode frequency
$\tilde{\omega}_{nlm}$ is given by
\begin{equation}
\tilde{\omega}_{nlm}^2=\omref^2+\lambda_{nlm}\ .
\label{eq18}
\end{equation}
Following the concept of quasi-degenerate perturbation theory, the coupling
of only two modes in the presence of only one flow component was
investigated by~\inlinecite{roth99}. They found that due to coupling of two
modes by any poloidal velocity field the absolute magnitudes of the
frequency shifts are equal for both couplers, but the sign is negative for
the mode with the lower frequency. In the case of two multiplets where pairs
of modes are coupling the frequency shifts are a function of the azimuthal
order $m$. The functional dependence on $m$ is given by the absolute
magnitude of the Wigner-$3j$ symbol as a function of $m$. It then follows
that the shifts within one multiplet are either always negative or always
positive. From this it follows for the case of the meridional circulation
that due to the dependence of the Wigner-$3j$ symbol on $m$ the
frequency shifts are always symmetric to the mode with $m=0$ which in general
is also shifted.

We would also like to point out that due to the second selection
rule~(\ref{eq16}) the flow kernel $K_s$ vanishes if $l'=l$, i.e. there are no
contributions to the diagonal entries on the matrix $Z$ from the matrix
elements $H_{n'n,l'l}(m)$. As the matrix elements $H_{n'n,l'l}(m)$ are
complex and $\mathbf{Z}$ is Hermitian, a change of the orientation of
$\vecb{u}_0$ does not change the frequency shifts.

\section{Results}

The result of our investigation are frequency shifts for the mode multiplets
with $0\le l\le 300$ and $1\le n\le 30$ calculated numerically by
setting up the full coupling matrix (\ref{eq8}), i.e. taking all possible couplings of modes into
account,
and evaluating the eigenvalues.

\begin{figure}
\includegraphics[width=9cm]{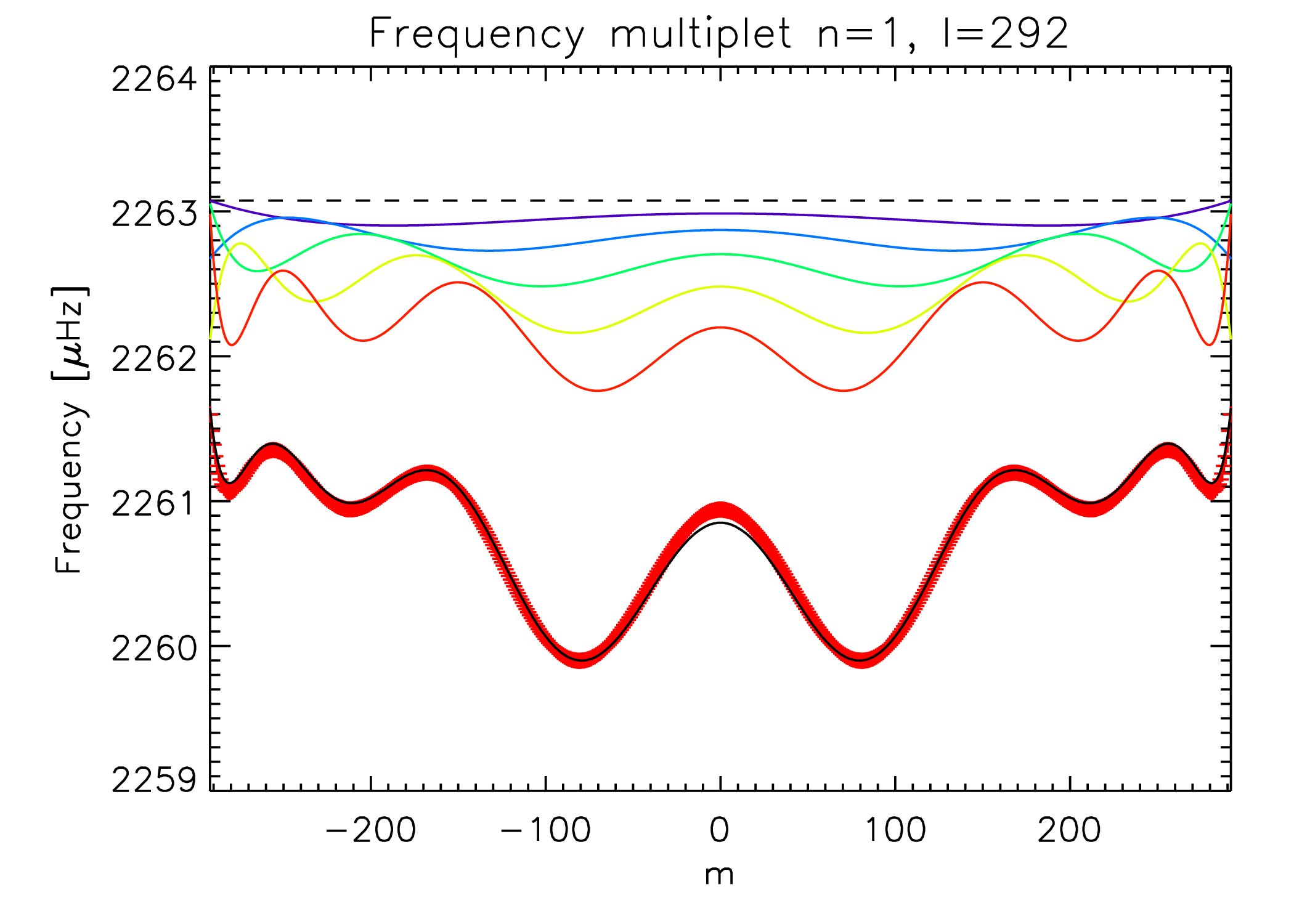}
\caption{Splittings of the p-mode multiplet $n=1$, $l=292$,
$\nu=2263.08$\thinspace $\mu$Hz as a function of azimuthal order $m$. The
splittings are caused by different meridional flow models: $s=2,\ n_c=1$
(purple), $s=3,\ n_c=1$ (blue), $s=4,\ n_c=1$ (green), $s=4,\ n_c=1$
(yellow), $s=4,\ n_c=1$ (red). The total effect of a sum of these five flow
models is displayed by the thick red line, the black line on top of this
gives the result for a sum of the five flow models but with $n_c=3$. The
unpertubed mode frequency is given by the dashed line. } \label{fig2}
\end{figure}

We first focus on the frequency shifts in the multiplets. According to the
properties of the Wigner-$3j$ symbols the shifts within a mulitplet are
symmetric to the mode with $m=0$. Figure~\ref{fig2} displays an example. The
frequency splitting is stronger for meridional circulation components with
higher degree $s$. In the example of Fig.~\ref{fig2} the shifts are in the
order of 1\thinspace $\mu$Hz for $s=6$. The overall effect of the
superposition of the meridional circulation components results in shifts in
the order of 3\thinspace $\mu$Hz away from the unperturbed frequency. The
difference between the summed-up effect of models with $n_c=1$ cell in depth
and models with $n_c=3$ cells in depth is in the order of 0.1\thinspace
$\mu$Hz.

\begin{figure}
\includegraphics[width=4cm]{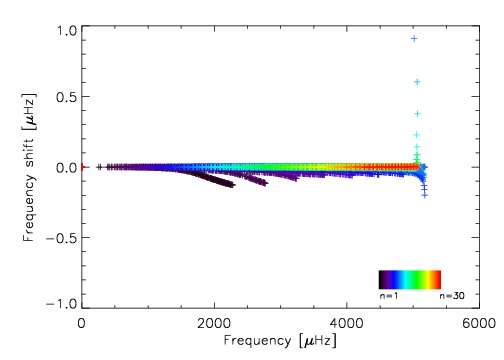}
\includegraphics[width=4cm]{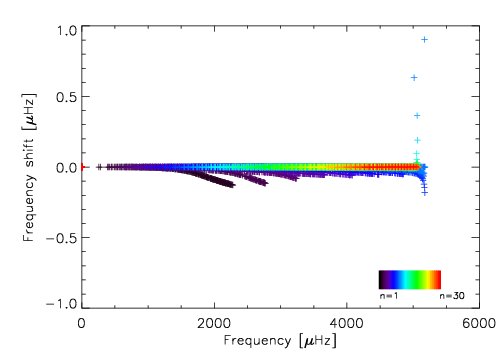}
\includegraphics[width=4cm]{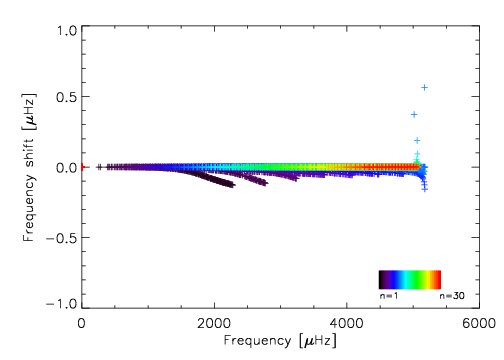}

\includegraphics[width=4cm]{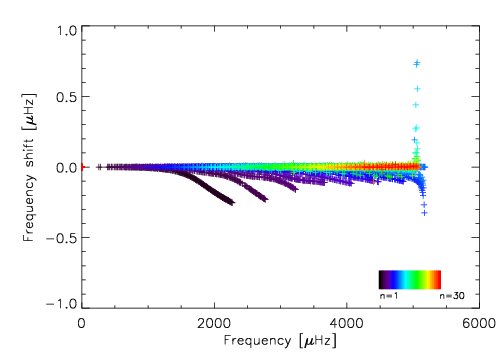}
\includegraphics[width=4cm]{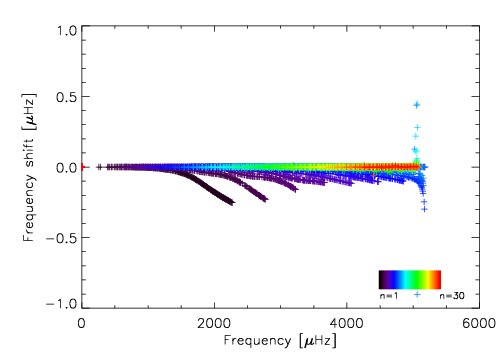}
\includegraphics[width=4cm]{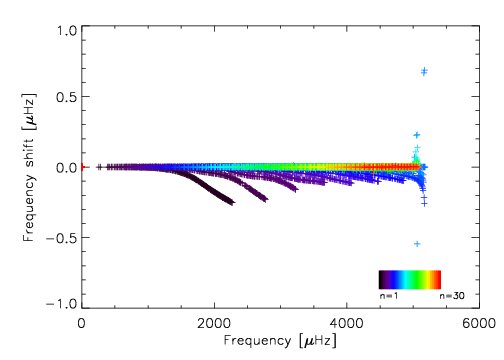}

\includegraphics[width=4cm]{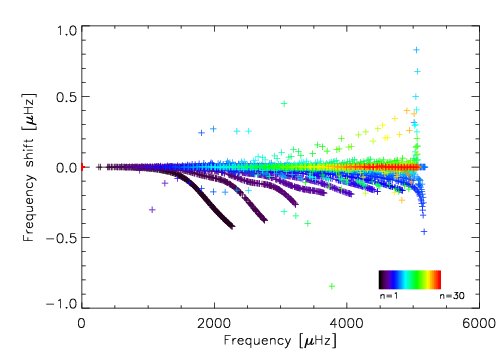}
\includegraphics[width=4cm]{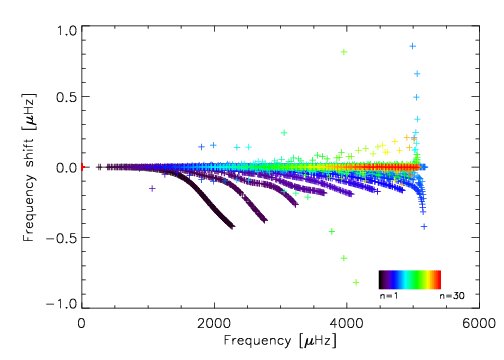}
\includegraphics[width=4cm]{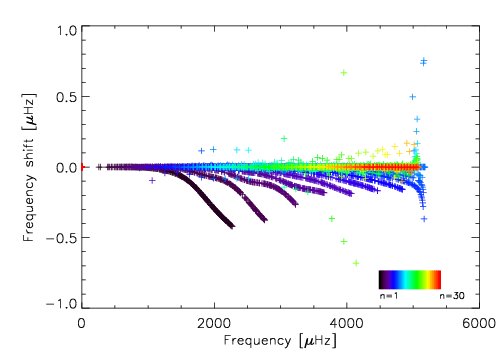}

\includegraphics[width=4cm]{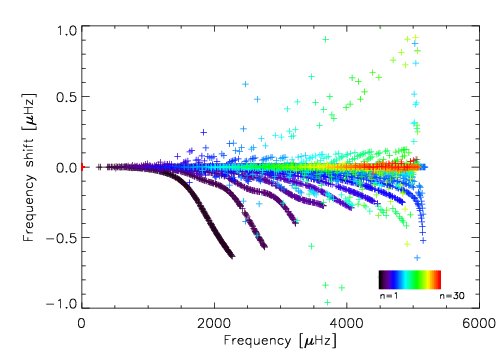}
\includegraphics[width=4cm]{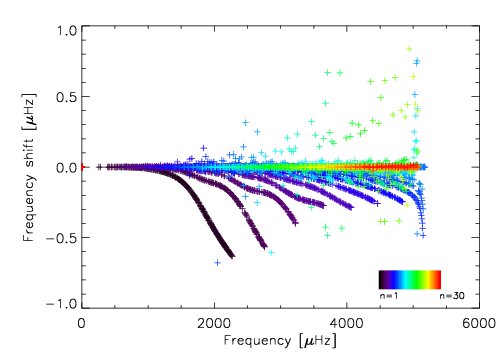}
\includegraphics[width=4cm]{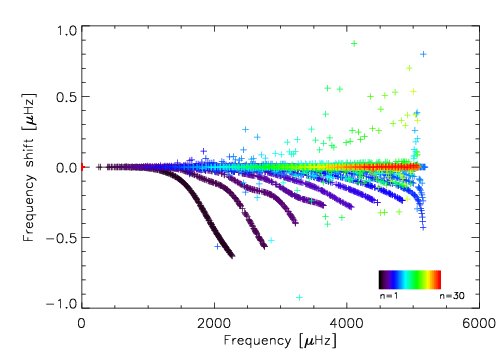}

\includegraphics[width=4cm]{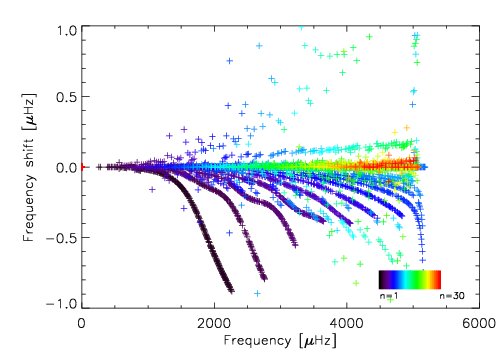}
\includegraphics[width=4cm]{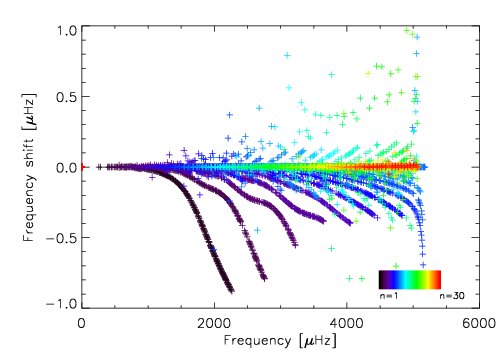}
\includegraphics[width=4cm]{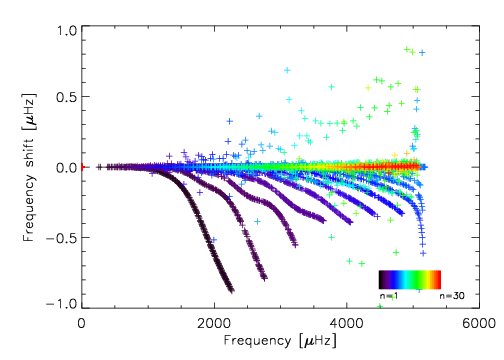}
\caption{The mean frequency shifts of the multiplets caused by the 15
different configurations of the meridional circulation (cf. Fig.~\ref{fig1})
as functions of the unpertubed mode frequencies. The color code gives the
radial order $n$ of the modes. The degree of the meridional circulation affecting
the modes changes from $s=2$ (top) to $s=6$ (bottom), the order $n_c$ of the
meridional circulation changes from 1 (left) to 3 (right).} \label{fig3}
\end{figure}

\begin{figure}
\includegraphics[width=9cm]{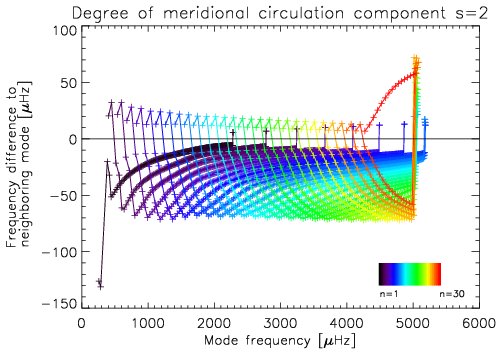}
\caption{The frequency difference to the next nearest mode
fulfilling the selection rule for the meridional circulation component with $s=2$. The
frequency difference is plotted versus mode frequency. The color code gives
the radial order of the modes.}
\label{fig4}
\end{figure}

\begin{figure}
\includegraphics[width=10cm]{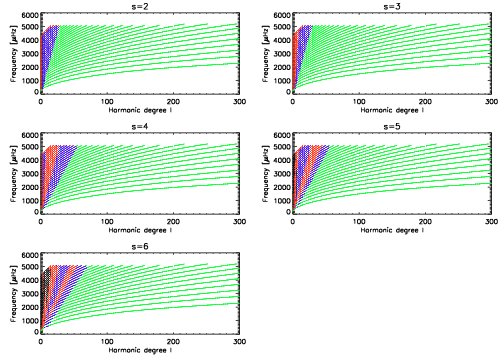}
\caption{Theoretical $l$-$\nu$ diagrams with colors indicating the origin of
the coupling partner with nearest frequency. Green: coupling partner from the
same ridge; blue: coupling partner from one ridge above; red: coupling
partner from one ridge below, black: coupling partner is more than one ridge
apart. The degrees $s$ of the meridional circulation components change from 2 to 6.}
\label{fig5}
\end{figure}

To give a better overview of the effect on all modes, Fig.~\ref{fig3}
displays the mean frequency shifts of the multiplets as a function of the
unperturbed mode frequencies. The frequency shifts were obtained for the
single meridional flow components.
One common result is a negative frequency shift of most modes. We find positive frequency
shifts only in a few multiplets. The
origin of this effect is given by the shape of the $l$-$\nu$ diagram.
According to quasi-degenerate perturbation theory the multiplet nearest in
frequency fulfilling the selection rules causes the strongest
shift~\cite{roth99}. As shown  for the case $s=2$ in Fig.~\ref{fig4} in most cases these nearest
neighboring modes have a higher frequency.
This is explained by the first selection rule and the curvature of the
ridges in the $l$-$\nu$ diagram. Because of the low degrees $s$ of the flow
components, the difference of the harmonic degrees of two coupling modes
$|l-l'|$ can not be larger than $s$. Possible coupling partners for a
particular mode can then only come from a very narrow region in $l$. As the
slopes of the ridges are steeper towards lower harmonic degrees, the
frequency spacing between the modes decreases along a ridge. In addition, the
frequency difference between two ridges increases with $l$. Therefore the
nearest neigboring mode fulfilling the selection rules usually lies on the
same ridge and has a higher frequency and a higher harmonic degree.
Figure~\ref{fig5} shows that in most of the investigated cases the
coupling partners lie on the same ridge.
Interestingly there is some structure visible in these plots. There are areas
where all coupling partners are coming from the same ridge or from one
ridge above or below. These areas are defined by the curvature of the ridges.
With higher degree $s$ of the meridional flow the coupling of modes from
different ridges becomes more
frequent. These plots indicate only the position of the nearest coupling
partner, because this partner causes the strongest effect. The effects of
coupling partners with greater distances in frequency are weaker.

Another common result is that the frequency shifts are strongest for the
modes with low radial order $n$ and with high harmonic degree $l$. This is
again explained by the location of the nearest neighbor in frequncy (cf.
Fig.~\ref{fig4}). The lower the radial order $n$ the smaller the frequency
difference to the next mode. Along a ridge the frequency difference to the
next mode decreases causing a stronger splitting of the multiplets. On
the ridges with high radial order, the frequency spacing between two
neigbouring modes is larger than on the ridges with low radial order.

These two common results apply in principle to all poloidal flows, not only
to the meridional circulation. Therefore we can conclude that the overall
effect of large-scale poloidal flows in the convection zone is a lowering of the mode
frequencies. This effect could contribute to reduce the discrepancy between
the theoretically calculated mode frequencies and the observed frequencies. A
similar result was already found by~\cite{zhugzhda94,stix98}.

Investigating the results in detail for the used meridional circulation
models we find that the frequency shifts are increasing from $s=2$ to $s=6$.
In the case of $s=2$, $n_c=1$ the total average of theses shifts for all modes affected by the
meridional circulation is 0.01\thinspace $\mu$Hz, whereas in the case $s=6$,
$n_c=1$ the magnitude of the average shift is 0.1\thinspace $\mu$Hz.

There is also a small difference noticeable in the frequency shift if the
number $n_c$ of cells changes. The difference of the average shifts
calculated for the models $s=2$, $n_c=1$ and $n_c=3$ is -0.002\thinspace
$\mu$Hz. Whereas the difference of the average shifts determined for the
models $s=6$, $n_c=1$ and $n_c=3$ is -0.03\thinspace $\mu$Hz.
However, the scatter in the frequency shifts is high. In the case of $s=2$,
$n_c=1$ the largest shift observed is 22.5\thinspace $\mu$Hz, and in the case
of n the case of $s=6$,
$n_c=1$ the largest shift observed is 79.4\thinspace $\mu$Hz.  Strong
shifts above 10\thinspace $\mu$Hz occur for the modes with high harmonic degree
$l$.

\section{Discussion}
In this paper we used simple models of the meridional circulation to
investigate their influences on the solar p-mode frequencies. The simplest
model consisted of one cell per hemisphere and depth with a maximum
horizontal flow velocity of 15\thinspace m/s on the solar surface. The most
complicated model we used had three cells per hemisphere and three revolution
cells in depth with a horizontal flow velocity of 15\thinspace m/s at the surface, too. We
performed numerical calculations based on quasi-degenerate perturbation
theory to obtain the frequency splittings of the solar p-modes due to this
meridional circulation models. We find that the meridional circulation lifts
the degeneracies of the multiplets. For the simplest model the shifts are on
average only 0.01\thinspace $\mu$Hz with a few shifts up to several $\mu$Hz.
Models with more cells per hemisphere affect the p-modes stronger. For the
model with three cells per hemisphere we find an average shift of
0.1\thinspace $\mu$Hz, with many shifts in the order of 1\thinspace $\mu$Hz.
In most cases the shifts are negative due to the fact that the next
neigboring mode in frequency dominates the shifting; and due to structure of
the $l$-$\nu$ diagram this nearest neighbor has usually a higher frequency
causing therefore negative shifts.

Comparing models with the same number of cells in latitude but different
number of cells in depth we find only tiny differences in the resulting
frequency splittings. In the case of $s=6$ the difference in the frequency
splitting is on average 0.03\thinspace $\mu$Hz. However a number of
differences in the order of 1\thinspace $\mu$Hz can occur.

On the Sun the meridional circulation consists probably of a superposition of
flow components with different numbers of cells in depth and latitude. In contrast to the
models used in our work, the amplitudes of these flow are in reality
likely to be very different and highly variable in time. Nevertheless, based on our results we are optimistic that
the meridional circulation on the Sun could leave an
observable signature in the p-mode frequencies. In order to detect the effect a
frequency resolution of at least 0.1\thinspace $\mu$Hz must be achieved.
This could be done by a several month long time series. In order to be able
to distinguish not only the various components of the meridional circulation
in the orders $s$ but also in the radial orders $n_c$, several years of data need
to be averaged to obtain the required precision in the frequencies.
As such long data sets exist it will be worth while to search for this effect
in the p-mode frequencies.

Compared to the splitting caused by the differential rotation the effect of meridional
circulation is small. But as the effect of the rotational splitting is odd it can be
separated from the even meridional splitting. However, e.g.
asphericities and the magnetic field might cause symmetrical
frequency shifts, too. Therefore, a lot of forward modelling will be
necessary before the effect of the meridional circulation can be disentangled from
these other effects. In this sense, one possible future extension of our work is the determination of the
frequency shifts due to more sophisticated models of the meridional
circulation, e.g. from three-dimensional numerical models. This might allow
tailoring inversion routines for estimating the meridional circulation in the Sun from global solar
oscillation frequencies.

\bibliographystyle{spr-mp-sola}

\bibliography{mrothlib.bib}

\end{article}

\end{document}